\documentclass[12pt,preprint]{aastex}
\usepackage{epsf}

\shorttitle{Cosmological Shock Waves}
\shortauthors{Kang {\it et al.~}}

\def\etal{{\it et al.}}

\def\ie{{\it i.e.,~}}

\def\kms{~{\rm km~s^{-1}}}
\def\cm3{~{\rm cm^{-3}}}

\def\Mpc{~h^{-1}{\rm Mpc}}
\def\kpc{~h^{-1}{\rm kpc}}

\begin{document}
\title{Cosmological Shock Waves in the Large Scale Structure
of the Universe: Non-gravitational Effects}

\author{Hyesung Kang\altaffilmark{1},
        Dongsu Ryu\altaffilmark{2},
        Renyue Cen\altaffilmark{3},
    and J. P. Ostriker\altaffilmark{3}}

\altaffiltext{1}
{Department of Earth Sciences, Pusan National University, Pusan 609-735,
Korea:\\ kang@uju.es.pusan.ac.kr} 
\altaffiltext{2}
{Department of Astronomy \& Space Science, Chungnam National University,
Daejeon 305-764, Korea:\\ ryu@canopus.cnu.ac.kr}
\altaffiltext{3}
{Department of Astrophysical Sciences, Princeton University, Princeton,
NJ 08544-1001, USA:\\ cen@astro.pricenton.edu, jpo@astro.pricenton.edu}

\begin{abstract}

Cosmological shock waves result from supersonic flow motions induced
by hierarchical clustering of nonlinear structures in the universe.
These shocks govern the nature of cosmic plasma through 
thermalization of gas and acceleration of nonthermal, cosmic-ray (CR)
particles.
We study the statistics and energetics of shocks formed in
cosmological simulations of a concordance $\Lambda$CDM universe,
with a special emphasis on the effects of non-gravitational
processes such as radiative cooling, photoionization/heating, and 
galactic superwind feedbacks.
Adopting an improved model for gas thermalization and CR acceleration
efficiencies based on nonlinear diffusive shock acceleration
calculations, we then estimate the gas thermal energy and the CR
energy dissipated at shocks through the history of the universe.
Since shocks can serve as sites for generation of vorticity,
we also examine the vorticity that should have been generated
mostly at curved shocks in cosmological simulations.
We find that the dynamics and energetics of shocks are governed
primarily by the gravity of matter, so other non-gravitational
processes do not affect significantly the global energy dissipation and
vorticity generation at cosmological shocks.
Our results reinforce scenarios in which the intracluster medium and
warm-hot intergalactic medium contain energetically significant
populations of nonthermal particles and turbulent flow motions.
 
\end{abstract}


\section{Introduction}

Astrophysical plasmas consist of both thermal particles and 
nonthermal, cosmic-ray (CR) particles that are closely coupled with
permeating magnetic fields and underlying turbulent flows.
In the interstellar medium (ISM) of our Galaxy, for example, 
an approximate energy equipartition among different components
seems to have been established,
\ie $\varepsilon_{therm} \sim \varepsilon_{CR} \sim \varepsilon_B
\sim \varepsilon _{turb} \sim 1\ {\rm eV\ cm^{-3}}$ \citep{long94}.
Understanding the complex network of physical interactions among 
these components constitutes one of fundamental problems in
astrophysics.

There is substantial observational evidence for the presence of
nonthermal particles and magnetic fields in the large scale structure
of the universe.
A fair fraction of X-ray clusters have been observed in diffuse 
radio synchrotron emission, indicating the presence of GeV CR
electrons and $\mu$G fields in the intracluster medium 
(ICM) \citep{gf00}.
Observations in EUV and hard X-ray have shown that some clusters
possess excess radiation compared to what is expected from the hot,
thermal X-ray emitting ICM, most likely produced by the inverse-Compton
scattering of cosmic background radiation (CBR) photons by CR electrons
\citep{fdfg99,bbk99,bbk00}.
Assuming energy equipartition between CR electrons and magnetic fields,
$\varepsilon_{CRe} \sim \varepsilon_B \sim 0.01-0.1 {\rm eV\ cm^{-3}} 
\sim 10^{-3}-10^{-2} \varepsilon_{therm}$ can be inferred 
in typical radio halos \citep{gf04}.
If some of those CR electrons have been energized at shocks and/or by
turbulence, the same process should have produced a greater CR proton
population.
Considering the ratio of proton to electron numbers, $K\sim 100$,
for Galactic CRs \citep{bk05}, one can expect 
$\varepsilon_{CRp} \sim 0.01-0.1 \varepsilon_{therm}$ in radio halos.
However, CR protons in the ICM have yet to be confirmed by the
observation of $\gamma$-ray photons produced by inelastic collisions
between CR protons and thermal protons \citep{rpsm03}.
Magnetic fields have been also directly observed with Faraday rotation
measure (RM).
In clusters of galaxies strong fields of a few $\mu$G strength extending
from core to 500 kpc or further were inferred from RM observations
\citep{ckb01,clar04}.
An {\it upper limit} of $\la \mu$G was imposed on the magnetic
field strength in filaments and sheets, based the observed limit of
the RMs of quasars outside clusters \citep{kron94,rkb98}.

Studies on turbulence and turbulent magnetic fields in the large scale
structure of the universe have been recently launched too.
XMM-Newton X-ray observations of the Coma cluster, which seems to be
in a post-merger stage, were analyzed in details to extract
clues on turbulence in the ICM \citep{sfmb04}.
By analyzing pressure fluctuations, it was shown that the turbulence
is likely subsonic and consistent with Kolmogoroff turbulence.
RM maps of clusters have been analyzed to find the power spectrum of
turbulent magnetic fields in a few clusters \citep{mgfg04,ve05}.
While \citet{mgfg04} reported a spectrum shallower than the Kolmogoroff
spectrum, \citet{ve05} argued that the spectrum could be consistent with
the Kolmogoroff spectrum if it is bended at a few kpc scale.
These studies suggest that as in the ISM, turbulence does exist in
the ICM and may constitute an energetically non-negligible component.

In galaxy cluster environments there are several possible sources of
CRs, magnetic fields, and turbulence:
jets from active galaxies \citep{kcld04,llfl06},
termination shocks of galactic winds
driven by supernova explosions \citep{volk99},
merger shocks \citep{sarazin99,gb03,fujita03},
structure formation shocks \citep{lw00,mjkr01,mrkj01}, 
and motions of subcluster clumps and galaxies \citep{subramanian06}.
All of them have a potential to inject a similar amount of energies, 
\ie $E \sim 10^{61}-10^{62}$ ergs into the ICM.
Here we focus on shock scenarios.

Astrophysical shocks are collisionless shocks that form in tenuous
cosmic plasmas via collective electromagnetic interactions between
gas particles and magnetic fields.
They play key roles in governing the nature of cosmic plasmas: \ie 
1) shocks convert a part of the kinetic energy of bulk flow motions
into thermal energy,
2) shocks accelerate CRs by diffusive shock acceleration (DSA)
\citep{bo78,be87,md01}, and amplify magnetic fields by streaming CRs
\citep{bell78,lucekbell00}, 
3) shocks generate magnetic fields via the Biermann battery
mechanism \citep{biermann50,kulsrude97} and the Weibel instability 
\citep{weibel59,medvedev06},
and 4) curved shocks generate vorticity and ensuing turbulent flows 
\citep{binney74,davies00}.

In \citet{rkhj03} (Paper I), the properties of {\it cosmological
shock waves} in the intergalactic medium (IGM) and the energy
dissipations into thermal and nonthermal components at those shocks
were studied in a high-resolution, adiabatic (non-radiative),
hydrodynamic simulation of a $\Lambda$CDM universe.
They found that internal shocks with low Mach numbers of $M \la 4$,
which formed in the hot, previously shocked gas inside nonlinear
structures, are  responsible for most of the shock energy dissipation.
Adopting a nonlinear DSA model for CR protons, it was shown that about
$1/2$ of the gas thermal energy dissipated at cosmological shocks
through the history of the universe could be stored as CRs.
In a recent study, \citet{pfrommeretal06} identified shocks and
analyzed the statistics in smoothed particle hydrodynamic (SPH)
simulations of a $\Lambda$CDM universe, and found that their results
are in good agreement with those of Paper I.
While internal shocks with lower Mach numbers are energetically
dominant, external accretions shocks with higher Mach numbers can
serve as possible acceleration sites for high energy cosmic rays
\citep{krj96,krb97,ostrowski02}.
It was shown that CR ions could be accelerated up to
$\sim Z \times 10^{19}eV$ at cosmological shocks, where $Z$ is
the charge of ions \citep{isma07}.

\citet{rkcd07} (Paper II) analyzed the distribution of vorticity,
which should have been generated mostly at cosmological shock waves,
in the same simulation of a $\Lambda$CDM universe as in
Paper I, and studied its implication on turbulence and turbulence dynamo.
Inside nonlinear structures, vorticity was found to be large enough
that the turn-over time, which is defined as the inverse of vorticity,
is shorter than the age of the universe.
Based on it \citet{rkcd07} argued that turbulence should have been
developed in those structures and estimated the strength of the magnetic
field grown by the turbulence.

In this paper, we study cosmological shock waves in a new set of
hydrodynamic simulations of large structure formation in a
concordance $\Lambda$CDM universe:
an adiabatic (non-radiative) simulation which is similar to that
considered in Paper I, and two additional simulations which include
various non-gravitational processes (see the next section for details). 
As in Papers I and II, the properties of cosmological shock waves are
analyzed, the energy dissipations to gas thermal energy and CR
energy are evaluated, and the vorticity distribution is analyzed.
We then compare the results for the three simulations to highlight
the {\it effects of non-gravitational processes} on the properties of
shocks and their roles on the cosmic plasmas in the large scale
structure of the universe.

Simulations are described in \S2.
The main results of shock identification and properties, energy
dissipations, and vorticity distribution are described in \S3,
\S 4, and \S 5, respectively.
Summary and discussion are followed in \S 6.

\section{Simulations}

The results reported here are based on the simulations previously
presented in \citet{co06}.
The simulations included radiative processes of heating/cooling, and
the two simulations with and without galactic superwind (GSW) feedbacks
were compared in that paper.
Here an additional adiabatic (non-radiative) simulation
with otherwise the same setup was performed.
Hereafter these three simulations are referred as ``Adiabatic'',
``NO GSW'', and ``GSW'' simulations, respectively.
Specifically, the WMAP1-normalized $\Lambda$CDM cosmology was employed
with the following parameters: 
$\Omega_{b}=0.048$, $\Omega_{m}=0.31$, $\Omega_{\Lambda}=0.69$,
$h \equiv H_0$/(100 km/s/Mpc) = 0.69, $\sigma_8 = 0.89$, and $n=0.97$. 
A cubic box of comoving size $85\Mpc$ was simulated using $1024^3$
grid zones for gas and gravity and $512^3$ particles for dark matter. 
It allows a uniform spatial resolution of $\Delta l = 83\kpc$.
In Papers I and II, an adiabatic simulation in a cubic box of comoving
size $100\Mpc$ with $1024^3$ grid zones and $512^3$ particles,
employing slightly different cosmological parameters, was used.
The simulations were performed using a PM/Eulerian
hydrodynamic cosmology code \citep{rokc93}.

Detailed descriptions for input physical ingredients such as
non-equilibrium ionization/cooling, photoionization/heating, 
star formation, and feedback processes can be found 
in earlier papers \citep{co03,co06}.
Feedbacks from star formation were treated in three forms:
ionizing UV photons, GSWs, and metal enrichment.
GSWs were meant to represent cumulative supernova explosions, and
modeled as outflows of several hundred $\kms$.
The input of GSW energy for a given amount of star formation was
determined by matching the outflow velocities computed for star-burst
galaxies in the simulation with those observed in the real world  
\citep{pebp02}\citep[see also][for details]{co06}.

Figure 1 shows the gas mass distribution in the gas density-temperature 
plane, $f_m(\rho_{gas}, T)$, and the gas mass fraction as a function of
gas temperature, $f_m(T)$, at $z=0$ for the three simulations.
The distributions are quite different, depending primarily on the
inclusion of radiative cooling and photoionization/heating.
GSW feedbacks increase the fraction of the WHIM with $10^5<T<10^7$K,
and at the same time affect the distribution of the warm/diffuse gas
with $T<10^5$.

\section{Properties of Cosmological Shock Waves}

We start to describe cosmological shocks by briefing the procedure
by which the shocks were identified in simulation data.
The details can be found in Paper I.
A zone was tagged as a {\it shock zone} currently experiencing shock
dissipation, whenever the following three criteria are met:
1) the gradients of gas temperature and entropy have the same sign,
2) the local flow is converging with ${\vec \nabla} \cdot {\vec v} < 0$,
and 3) $|\Delta \log T| \ge 0.11$ corresponding to the temperature 
jump of a shock with $M \ge 1.3$.
Typically a shock is represented by a jump spread over $2-3$ tagged
zones.
Hence, a {\it shock center} was identified within the tagged zones,
where ${\vec \nabla} \cdot {\vec v}$ is minimum, and this center
was labeled as part of a shock surface.
The Mach number of the shock center, $M$, was calculated from the
temperature jump across the entire shock zones.
Finally to avoid confusion from complex flow patterns and shock surface
topologies associated with very weak shocks, only those portions of
shock surfaces with $M\ge1.5$ were kept and used for the analysis of
shocks properties.

Figure 2 shows the locations of identified shocks in a two-dimensional
slice at $z=0$ in the GSW simulation. 
The locations are color-coded according to shock speed.
As shown before in Paper I, {\it external} accretion shocks encompass
nonlinear structures and reveal, in addition to cluster complexes,
rich topology of filamentary and sheet-like structures in the large
scale structure.
Inside the  nonlinear structures, there exist complex networks of
{\it internal} shocks that form by infall of previously shocked gas
to filaments and knots and during subclump mergers, as well as by
chaotic flow motions.
The shock heated gas around clusters extends out to $\sim 5 \Mpc$,
much further out than the observed X-ray emitting volume.

In the GSW simulation, with  several hundred $\kms$ for outflows,
the GSW feedbacks affected most greatly the gas around groups of
galaxies, while the impact on clusters with $kT \ga 1$ keV was minimal.
In Figure 3 we compare shock locations in a region around two groups
with $kT \sim 0.2-0.3$ keV in the three simulations.
It demonstrates that GSW feedbacks pushed the hot gas out of
groups with typical velocities of $\sim 100 \kms$ (green points). 
In fact the prominent green balloons of shock surfaces around 
groups in Figure 2 are due to GSW feedbacks
\citep[see also Figure 4 of][]{co06}.

In the left panels of Figure 4 we compare the surface area of
identified shocks, normalized by the volume of the simulation box, per
logarithmic Mach number interval, $dS(M)/d\log M$ (top), and per
logarithmic shock speed interval, $dS(V_s)/d\log V_s$ (bottom),
at $z=0$ in the three simulations.
Here $S$ and $V_s$ are given in units of $(h^{-1}{\rm Mpc})^{-1}$
and $\kms$. 
The quantity $S$ provides a measure of shock frequency or the inverse
of the mean comoving distance between shock surfaces. 
The distributions of $dS(M)/d\log M$ for the NO GSW and GSW simulations
are similar, while that for the Adiabatic simulation is different from
the other two.
This is mainly because the gas temperature outside nonlinear structures
is lower without photoionization/heating in the Adiabatic simulation. 
As a result, external accretion shocks tend to have higher Mach number
due to colder preshock gas. 
The distribution of $dS(V_s)/d\log V_s$, on the other hand, is similar
for all three simulations for $V_s> 15 \kms$.
For $V_s<15 \kms$, however, there are more shocks in the Adiabatic
simulation (black points in Figure 3).
Again this is because in the Adiabatic simulation the gas temperature
is colder in void regions, and so even shocks with low speeds of
$V_s<15 \kms$ were identified in these regions.
The GSW simulation shows slightly more shocks than the NO GSW simulation
around $V_s \sim 100\kms$, because GSW
feedbacks created balloon-shaped surfaces of shocks with typically
those speeds (green points in Figure 3). 

For identified shocks, we calculated the incident shock kinetic energy
flux, $F_{\phi}= (1/2) \rho_1 V_{s}^3$, where $\rho_1$ is the preshock
gas density.
We then calculated the kinetic energy flux through shock surfaces,
normalized by the volume of the simulation box, per logarithmic Mach
number interval, $dF_{\phi}(M)/d\log M$, and per logarithmic shock
speed interval, $dF_{\phi}(V_s)/d\log V_s$.
In the right panels of Figure 4, we compare the flux at $z=0$ in
the three simulations.
Once again, there are noticeable differences in $dF_{\phi}(M)/d\log M$
between the Adiabatic simulation and the other two simulations, which
can be interpreted as the result of ignoring photoionization/heating
in the gas outside nonlinear structures in the Adiabatic simulation. 
GSW feedbacks enhance only slightly the shock kinetic energy flux for 
$V_s \sim 100 - 300\kms$, as can be seen in the plot of
$dF_{\phi}(V_s)/d\log V_s$.
Yet, the total amount of the energy flux is expected to be quite similar
for all three simulations.
This implies that the overall energy dissipation at cosmological shocks
is governed mainly by the gravity of matter, and that the inclusion
of various non-gravitational processes such as radiative cooling,
photoionization/heating, and GSW feedbacks have rather minor, local
effects. 

We note that a {\it temperature floor} of $T_{\rm floor} = T_{\rm CBR}$
was used for the three simulations in this work, while
$T_{\rm floor} = 10^4$ K was set in paper I.
It was because in Paper I only an adiabatic simulation was considered
and the $10^4$ K temperature floor was enforced to mimic the effect of
photoionization/heating on the IGM.
However we found that when the same temperature floor is enforced, the
statistics of the current Adiabatic simulation agree excellently with
those of Paper I.
Specifically, the shock frequency and kinetic energy flux,
$dS(M)/d\log M$ and $dF_{\phi}(M)/d\log M$, for weak shocks with
$1.5 \le M \la 3$ are a bit higher in the current Adiabatic simulation,
because of higher spatial resolution.
But the total kinetic energy flux through shock surfaces,
$F_{\phi}(M>1.5)$, agrees within a few percent.
On the other hand, In Paper I we were able to reasonably distinguish
external and internal shocks according to the preshock temperature,
\ie external shocks if $T_1 \le T_{\rm floor}$ and internal shocks
if $T_1>T_{\rm floor}$.
We no longer made such distinction in this work, since the preshock 
temperature alone cannot tell us whether the preshock gas is 
inside nonlinear structures or not in the simulations with
radiative cooling. 

\section{Energy dissipation by Cosmological Shock Waves}

The CR injection and acceleration rates at shocks depend in general 
upon the shock Mach number, field obliquity angle, and the strength
of the Alfv\'en turbulence responsible for scattering.
At quasi-parallel shocks, in which the mean magnetic field is parallel
to the shock normal direction, small anisotropy in the particle
velocity distribution in the local fluid frame causes some particles
in the high energy tail of the Maxwellian distribution to stream
upstream \citep{gbse92}.
The streaming motions of the high energy particles against the
background fluid generate strong MHD Alfv\'en waves upstream of
the shock, which in turn scatter particles and amplify magnetic
fields \citep{bell78,lucekbell00}.
The scattered particles can then be accelerated further to higher
energies via Fermi first order process \citep{md01}.
These processes, \ie leakage of suprathermal particles into CRs,
self-excitation of Alfv\'en waves, amplification of magnetic fields,
and further acceleration of CRs, are all integral parts of collisionless
shock formation in astrophysical plasmas.
It was shown that at strong quasi-parallel shocks, $10^{-4} - 10^{-3}$
of the incoming particles can be injected into the CR population, up
to 60\% of the shock kinetic energy can be transferred into CR ions,
and at the same time substantial nonlinear feedbacks are exerted to
the underlying flow \citep{ber95,kj05}. 

At perpendicular shocks with weakly perturbed magnetic fields, on the
other hand, particles gain energy mainly by drifting along the shock
surface in the ${\vec v} \times {\vec B}$ electric field.
Such drift acceleration can be much more efficient than the acceleration
at parallel shocks \citep{jokipii87,krb97,ostrowski02}. 
But the particle injection into the acceleration process is expected to
be inefficient at perpendicular shocks, since the transport of particles
normal to the average field direction is suppressed \citep{ellisonetal95}.
However, \citet{giacalone05} showed that the injection problem at
perpendicular shocks can be alleviated substantially in the presence
of fully turbulent fields owing to field line meandering.

As in Paper I, the gas thermalization and CR acceleration
efficiencies are defined as $\delta(M) \equiv F_{th}/F_{\phi}$ and
$\eta(M) \equiv F_{CR}/F_{\phi}$, respectively,
where $F_{th}$ is the thermal energy flux generated
and $F_{CR}$ is the CR energy flux accelerated at shocks.
We note that for gasdynamical shocks without CRs, the gas
thermalization efficiency can be calculated from the Rankine-Hugoniot
jump condition, as follows:
\begin{equation}
\delta_0(M) = \left[e_{th,2} - e_{th,1} \left({\rho_2\over\rho_1}
\right)^{\gamma}\right]v_2 \Bigg/ \left({1\over2}\rho_1 v_1^2\right),
\end{equation}
where the subscripts 1 and 2 stand for preshock and postshock regions, 
respectively.
The second term inside the brackets subtracts the effect of adiabatic
compression occurred at a shock too, not just the thermal energy flux
entering the shock, namely, $e_{th,1} v_1$.

At CR modified shocks, however, the gas thermalization efficiency
can be much smaller than $\delta_0(M)$ for strong shocks 
with large $M$, since a significant fraction of the shock
kinetic energy can be transferred to CRs.
The gas thermalization and CR acceleration efficiencies were estimated
using the results of DSA simulations of quasi-parallel shocks with
Bohm diffusion coefficient, self-consistent treatments of thermal
leakage injection, and Alfv\'en wave propagation \citep{kj07}.
The simulations were started with purely gasdynamical shocks in
one-dimensional, plane-parallel geometry, and CR acceleration was
followed by solving the diffusion-convection equation explicitly
with very high resolution.
Shocks with $V_s=150-4500$ km s$^{-1}$ propagating into media of
$T_1 = 10^4 - 10^6$ K were considered.
After a quick initial adjustment, 
the postshock states reach time asymptotic values  
and the CR modified shocks evolve in an approximately self-similar way with 
the shock structure broadening linearly with time
\citep[refer][for details]{kj07}.
Given this self-similar nature of CR modified shocks, we calculated
time asymptotic values of $\delta(M)$ and $\eta(M)$ as the ratios of
increases in the gas thermal and CR energies at shocks to the kinetic
energy passed through the shocks at the termination time of the
DSA simulations.
As in Eq. (1), the increase of energies due to adiabatic compression
was subtracted.

Figure 5 shows $\delta(M)$ and $\eta(M)$ estimated from DSA simulations and
their fittings for the cases with and without a preexisting CR component.
The fitting formulae are given in Appendix A.
Without a preexisting CR component, gas thermalization is more
efficient than CR acceleration at shocks with $M \la 5$.
However, it is likely that weak internal shocks propagate through
the IGM that contains CRs accelerated previously at earlier shocks.
In that case, shocks with preexisting CRs need to be considered. 
Since the presence of preexisting CRs is equivalent to a higher
injection rate, CR acceleration is more efficient in that case,
especially at shocks with $M \la 5$ \citep{kj03}.
In the bottom panel the efficiencies for shocks with
$P_{CR}/P_{g} \sim 0.3$ in the preshock region are shown.
For comparison, $\delta_0(M)$ for shocks without CRs is also drawn. 
Both $\delta(M)$ and $\eta(M)$ increase with Mach number, but
$\eta(M)$ asymptotes to $\sim 0.55$ while $\delta(M)$ to $\sim 0.30$
for strong shocks with $M \ga 30$.
So about twice more energy goes into CRs, compared to for gas heating,
at strong shocks.
 
The efficiencies for the case without a preexisting CR component in
the upper panel of Figure 5 can be directly compared with the same
quantities presented in Figure 6 of Paper I.
In Paper I, however, the gas thermalization efficiency was not calculated 
explicitly from DSA simulations, and hence $\delta_0(M)$ for gasdynamic
shocks was used.
It represents gas thermalization reasonably well for weak
shocks with $M \la 2.5$, but overestimates gas thermalization for
stronger CR modified shocks.
Our new estimate for $\eta(M)$ is close to that in Paper I, but a bit 
smaller, especially for shocks with $M \la 30$.
This is because inclusion of Alfv\'en wave drift and dissipation in the
shock precursor reduces the effective velocity change experienced by
CRs in the new DSA simulations of \citet{kj07}. 

A note of caution for $\eta(M)$ should be in order.
As outlined above, CR injection is less efficient  and so the CR
acceleration efficiency would be lower at perpendicular shocks,
compared to at quasi-parallel shocks.
CR injection and acceleration at oblique shocks are not well understood 
quantitatively.
And the magnetic field directions at cosmological shocks are not known.
Considering these and other uncertainties involved in the adopted DSA
model, we did not attempt to make further improvements in estimating
$\delta(M)$ and $\eta(M)$ at general oblique shocks.
But we expect that an estimate at realistic shocks with chaotic magnetic
fields and random shock obliquity angles would give reduced values,
rather than increased values, for $\eta(M)$.
So $\eta(M)$ given in Figure 5 may be regarded as {\it upper limits}.

By adopting the efficiencies in Figures 5, we calculated the thermal
and CR energy fluxes dissipated at cosmological shocks,
$dF_{th}(M)/d\log M$, $dF_{th}(V_s)/d\log V_s$, $dF_{CR}(M)/d\log M$
and $dF_{CR}(V_s)/d\log V_s$, using $F_{th}= F_{\phi}\delta(M)$ and
$F_{CR}=F_{\phi}\eta(M)$, in the same way we calculated
$dF_{\phi}(M)/d\log M$ and $dF_{\phi}(V_s)/d\log V_s$ in
the previous section. 
We then integrated  from $z=5$ to $z=0$ the shock kinetic energy
passed and the thermal and CR energies dissipated through shock
surfaces as follows:
\begin{equation}
{{d Y_i(X)} \over {d \log X}} = { 1 \over {\cal E}_{th,0}}
\int_{z=5}^{z=0} {{d F_i[X,z(t)]} \over {d \log X}} dt,
\end{equation}
where the subscript $i \equiv \phi,~ th,~{\rm or}~CR$ stands for
the kinetic, thermal, or CR energies fluxes, the variable $X$ is
either $M$ or $V_s$, and ${\cal E}_{th,0}$ is the total gas thermal
energy  at $z=0$ inside the simulation box normalized by its volume.

Figure 6 shows the resulting ${d Y_i(M)}/{d \log M}$ and
${d Y_i(V_s)}/{d \log V_s}$ and their cumulative distributions,
$Y_i(>M)$ and $Y_i(>V_s)$, for the GSW simulation.
Weak shocks with $ M \la 4$ or fast shocks with $V_s\ga 500 \kms$ are
responsible most for shock dissipations, as already noted in Paper I.
While the thermal energy generation peaks at shocks in the range
$1.5 \la M \la 3$, the CR energy peaks in the range
$2,5 \la M \la 4$ if no preexisting CRs are included or in the range
$1.5 \la M \la 3$ if preexisting CRs of $P_{CR}/P_g \sim 0.3$ in the
preshock region are included.
With our adopted efficiencies, the total CR energy accelerated and
the total gas thermal energy dissipated at cosmological shocks
throughout the history of the universe are compared as
$Y_{CR}(M \ge 1.5) \sim 0.5 Y_{th}(M \ge 1.5)$, when no preexisting
CRs are present.
With preexisting CRs in the preshock region,
the CR acceleration becomes more efficient, so
$Y_{CR}(M \ge 1.5) \sim 1.7 Y_{th}(M \ge 1.5)$,
\ie the total CR energy accelerated at cosmological shocks is
estimated to be 1.7 times the total gas thermal energy dissipated.
We note here again that these are not meant to be very
accurate estimates of the CR energy in the IGM, considering the
difficulty of modeling shocks as well as the uncertainties
in the DSA model itself.
However, they imply that the IGM and the WHIM, which are bounded by 
strong external shocks with high $M$ and filled with weak internal shocks
with low $M$, could contain a dynamically significant CR population.

\section{Vorticity Generation at Cosmological Shock Waves}

Cosmological shocks formed in the large scale structure of the
universe are by nature curved shocks, accompanying complex,
often chaotic flow patterns.
It is well known that vorticity, 
${\vec \omega}= \nabla \times {\vec v}$, is generated
at such curved oblique shocks \citep{binney74,davies00}.
In Paper II, the generation of vorticity behind cosmological shocks
and turbulence dynamo of magnetic fields in the IGM were studied in
an adiabatic $\Lambda$CDM simulation. 
In this study we analyzed the distribution of vorticity
in the three simulations to assess quantitatively the effects of
non-gravitational processes.
Here we present the magnitude of vorticity with the vorticity
parameter
\begin{equation}
\tau({\vec r},z) \equiv t_{\rm age}(z) \omega({\vec r},z) =
{t_{\rm age}(z) \over t_{\rm eddy}({\vec r},z)},
\end{equation}
where $t_{\rm age}(z)$ is the age of the universe at redshift $z$.
With $t_{\rm eddy} = 1/\omega$ interpreted as local eddy turnover
time, $\tau$ represents the number of local eddy turnovers in the
age of the universe.
So if $\tau \gg 1$, we expect that turbulence has been fully
developed after many turnovers. 

Figure 7 shows fluid quantities and shock locations in a
two-dimensional slice of $(21.25 \Mpc)^2$, delineated by a solid box
in Figure 2, at $z=0$ in the GSW simulations. 
The region contains two clusters with $kT \sim 1 - 2$ keV in
the process of merging. 
Bottom right panel shows that vorticity increases sharply at shocks.
The postshock gas has a larger amount of vorticity than the preshock
gas, indicating that most, if not all, of the vorticity in the
simulation was produced at shocks.

Figure 8 shows the gas mass distribution in the gas density-vorticity
parameter plane, $f_m(\rho_{gas},\tau)$, (upper panel) and the gas mass
fraction per logarithmic $\tau$ interval, $df_m(\tau)/d\log \tau$,
(bottom panel) for the three simulations.
The most noticeable point in the upper panel is that vorticity is higher
at the highest density regions with
$\tilde \rho \equiv \rho_{gas}/\langle \rho_{gas} \rangle \ga 10^3$
in the NO GSW and GSW simulations than in the Adiabatic simulation.
This is due to the additional flow motions induced by cooling.
Inclusion of GSW feedbacks, on the other hand, does not alter
significantly the overall distribution in the gas density-vorticity
parameter plane.
The bottom panel indicates that cooling increased the mass fraction
with large vorticity $\tau \ga 10$, while reduced the mass fraction
with $1 \la \tau \la 10$.
GSW feedbacks increased slightly the mass fraction with
$1\la \tau \la10$, which corresponds to the gas in the regions
outskirts of groups that expand further out due to GSWs
(\ie balloons around groups).
But overall we conclude that the non-gravitational processes
considered in this paper have limited effects on vorticity
in the large scale structure of the universe.

We note that the highest density regions in the NO GSW and GSW
simulations have $\tau \sim 30$ on average.
As described in details in Paper II, such values of $\tau$ imply that local
eddies have turned over many times in the age of the universe, so
that the ICM gas there has had enough time to develop
magnetohydrodynamic (MHD) turbulence.
So in those regions, magnetic fields should have grown to have
the energy approaching to the turbulent energy.
On the other hand, the gas with $1 \la \tilde \rho \la 10^3$,
mostly in filamentary and sheet-like structures, has $0.1 \la \tau \la 10$.
MHD turbulence should not have been fully developed there
and turbulence growth of magnetic fields would be small.
Finally in the low density void regions with $\tilde \rho \la 1$,
vorticity is negligible with $\tau \la 0.1$ on average, as expected.

\section{Summary}

We identified cosmological shock waves and studied their roles on
cosmic plasmas in three cosmological N-body/hydro\-dynamic simulations
for a concordance
$\Lambda$CDM universe in a cubic box of comoving size $85\Mpc$: 
1) adiabatic simulation (Adiabatic), 2) simulation with
radiative cooling and photoionization/heating (NO GSW), and 3) same as
the second simulation but also with galactic superwind feedbacks (GSW).
The statistics and energetics of shocks in the adiabatic simulation
are in an excellent agreement with those of Paper I where an adiabatic
simulation with slightly different cosmological parameters in
a cubic box of comoving size $100\Mpc$ was analyzed.

Photoionization/heating raised the gas temperature outside nonlinear
structures in the NO GSW and GSW simulations.
As a result, the number of identified shocks and their Mach numbers in
the NO GSW and GSW simulations were different from those in the Adiabatic
simulation.
GSW feedbacks pushed out gas most noticeably around groups, creating
balloon-shaped surfaces of shocks with speed $V_s \sim 100 \kms$ in
the GSW simulation.
However, those have minor effects on shock energetics.
The total kinetic energy passed through shock surfaces throughout the
history of the universe is very similar for all three simulations.
So we conclude that the energetics of cosmological shocks was governed
mostly by the gravity of matter, and the effects non-gravitational
processes, such as radiative cooling, photoionization/heating, and GSW
feedbacks, were rather minor and local.

We estimated both the {\it improved} gas
thermalization efficiency, $\delta(M)$, and CR acceleration efficiency,
$\eta(M)$, as a function shock Mach number, from nonlinear diffusive
shock simulations for quasi-parallel shocks that
assumed Bohm diffusion for CR protons and 
incorporated self-consistent treatments of
thermal leakage injection and Alfv\'en wave propagation \citep{kj07}.
The cases without and with a preexisting CR component of
$P_{CR}/P_g \sim 0.3$ in the preshock region were considered.
At strong shocks, both the injection and acceleration of CRs are
very efficient, and so the presence of a preexisting CR component
is not important.
At shocks with with $M \ga 30$, about 55 \% of the shock kinetic energy
goes into CRs, while about 30 \% becomes the thermal energy.
At weak shocks, on the other hand, without a preexisting CR component,
the gas thermalization is more efficient than the CR acceleration.
But the presence of a preexisting CR component is critical at weak shocks,
since it is equivalent to a higher injection rate and the CR acceleration
becomes more efficient with it.
As a result, $\eta(M)$ is higher than $\delta(M)$ even at shocks
with $M \la 5$.
However, at perpendicular shocks, the CR injection is suppressed, and
so the CR acceleration could be less efficient than at parallel shocks. 
Thus our CR shock acceleration efficiency should be regarded as
an upper limit.

With the adopted efficiencies, the total CR energy accelerated at
cosmological shocks throughout the history of the universe is estimated
to be $Y_{CR}(M \ge 1.5) \sim 0.5\ Y_{th}(M \ge 1.5)$, \ie 1/2 of the
total gas thermal energy dissipated, when no preexisting CRs are present.
With a preexisting CR component of $P_{CR}/P_g \sim 0.3$ in the preshock
region, $Y_{CR}(M \ge 1.5) \sim 1.7\ Y_{th}(M \ge 1.5)$, \ie the total
CR energy accelerated is estimate to be 1.7 times the total gas thermal
energy dissipated.
Although these are not meant to be very accurate estimates of the CR
energy in the ICM, they imply that the ICM could contain a dynamically
significant  CR population.

We also examined the distribution of vorticity inside the simulation box,
which should have been generated mostly at curved cosmological shocks.
In the ICM, the eddy turn-over time, $t_{eddy} = 1/\omega$, is about 1/30
of the age of the universe, \ie $\tau \equiv t_{age}/t_{eddy} \sim 30$.
In filamentary and sheet-like structures, $\tau \sim 0.1 -10$, while
$\tau \la 0.1$ in void regions.
Radiative cooling increased the fraction of gas mass with large vorticity
$\tau \ga 10$, while reduced the mass fraction with $1 \la \tau \la 10$.
GSW feedbacks increased slightly the mass fraction with $1\la \tau \la10$.
Although the effects of these non-gravitation effects are not negligible,
the overall distribution of vorticity are similar for the three simulations.
So we conclude that the non-gravitational processes considered in this
paper do not affect significantly the vorticity in the large scale
structure of the universe.

\acknowledgements
HK was supported in part by KOSEF through Astrophysical Research
Center for the Structure and Evolution of Cosmos (ARCSEC).
DR was supported in part by a Korea Research Foundation grant
(KRF-2004-015-C00213).
RC was supported in part by NASA grant NNG05GK10G and NSF grant
AST-0507521.
The work of HK and DR was also supported in part by Korea
Foundation for International Cooperation of Science \& Technology
(KICOS) through the Cavendish-KAIST Research Cooperation Center.

\appendix

\section{Fitting Formulae for $\delta(M)$ and $\eta(M)$}

The gas thermalization efficiency, $\delta(M)$, and the CR acceleration
efficiency, $\eta(M)$, for the case without a preexisting CR component
(in upper panel of Figure 5) are fitted as follows:\\
for $M \le 2$
\begin{equation}
\delta(M) = 0.92\ \delta_0
\end{equation}
\begin{equation}
\eta(M)=  1.96\times10^{-3} (M^2-1)
\end{equation}
for $M > 2$
\begin{equation}
\delta(M)= \sum_{n=0}^4 a_n {{(M-1)^n} \over M^4}
\end{equation}
\begin{equation}
a_0 = -4.25,~~~a_1 = 6.42,~~~a_2 = -1.34,~~~a_3 = 1.26,~~~a_4 = 0.275
\end{equation}
\begin{equation}
\eta(M)= \sum_{n=0}^4 b_n {{(M-1)^n} \over M^4}
\end{equation}
\begin{equation}
b_0 = 5.46,~~~b_1 = -9.78,~~~b_2 = 4.17,~~~b_3 =- 0.334,~~~b_4 = 0.570
\end{equation}

The efficiencies for the case with a preexisting CR component (in bottom
panel of Figure 5) are fitted as follows:\\
for $M \le 1.5$
\begin{equation}
\delta(M) = 0.90\ \delta_0
\end{equation}
\begin{equation}
\eta(M) = 1.025\ \delta_0
\end{equation}
for $M > 1.5$
\begin{equation}
\delta(M)= \sum_{n=0}^4 a_n {{(M-1)^n} \over M^4}
\end{equation}
\begin{equation}
a_0 = -0.287,~~~a_1 = 0.837,~~~a_2 = -0.0467,~~~a_3 = 0.713,~~~a_4 = 0.289
\end{equation}
\begin{equation}
\eta(M)= \sum_{n=0}^4 b_n {{(M-1)^n} \over M^4}
\end{equation}
\begin{equation}
b_0 = 0.240,~~~b_1 = -1.56,~~~b_2 = 2.80,~~~b_3 = 0.512,~~~b_4 = 0.557
\end{equation}

Here $\delta_0(M)$ is the gas thermalization efficiency at shocks without
CRs, which was calculated from the Rankine-Hugoniot jump condition,
(black solid line in Figure 5):
\begin{equation}
\delta_0 (M) = {2 \over \gamma (\gamma-1) M^2 R} \left[
{2\gamma M^2 - (\gamma-1) \over (\gamma+1)} - R^\gamma \right]
\end{equation}
\begin{equation}
R \equiv {\rho_2 \over \rho_1} = { {\gamma+1} \over {\gamma -1 + 2/M^2}}
\end{equation}

\clearpage

\begin{figure}
\vspace{-1cm}
\centerline{\epsfxsize=18cm\epsfbox{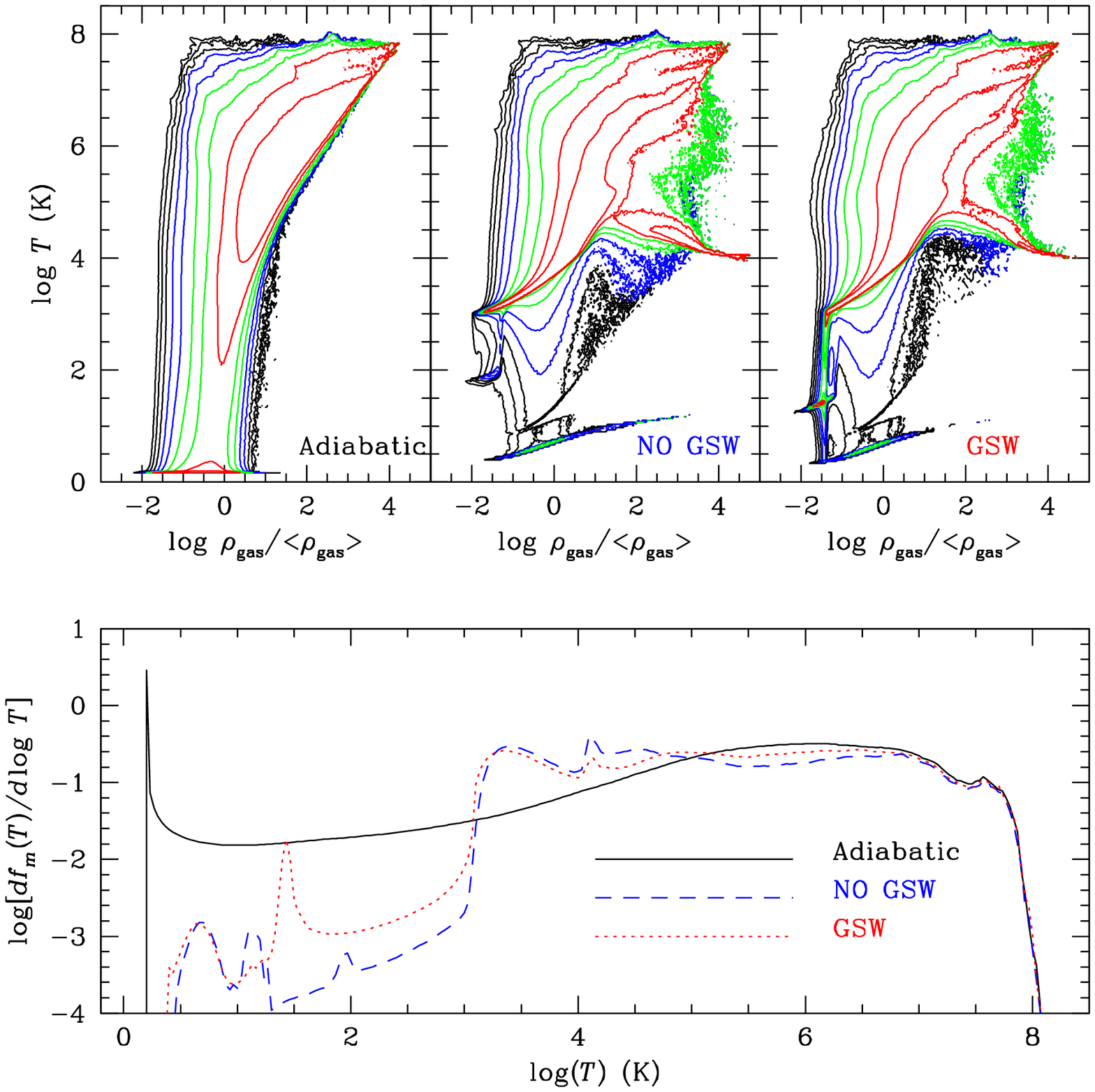}}
\figcaption{ {\it Top panels:} Gas mass distribution in the gas
density-temperature plane at $z=0$ for the Adiabatic, NO GSW, and
GSW simulations.
{\it Bottom panel:} Gas mass fraction as a function of gas
temperature at $z=0$ for the three simulations.}
\end{figure}

\clearpage

\begin{figure}
\vspace{-1cm}
\centerline{\epsfxsize=18cm\epsfbox{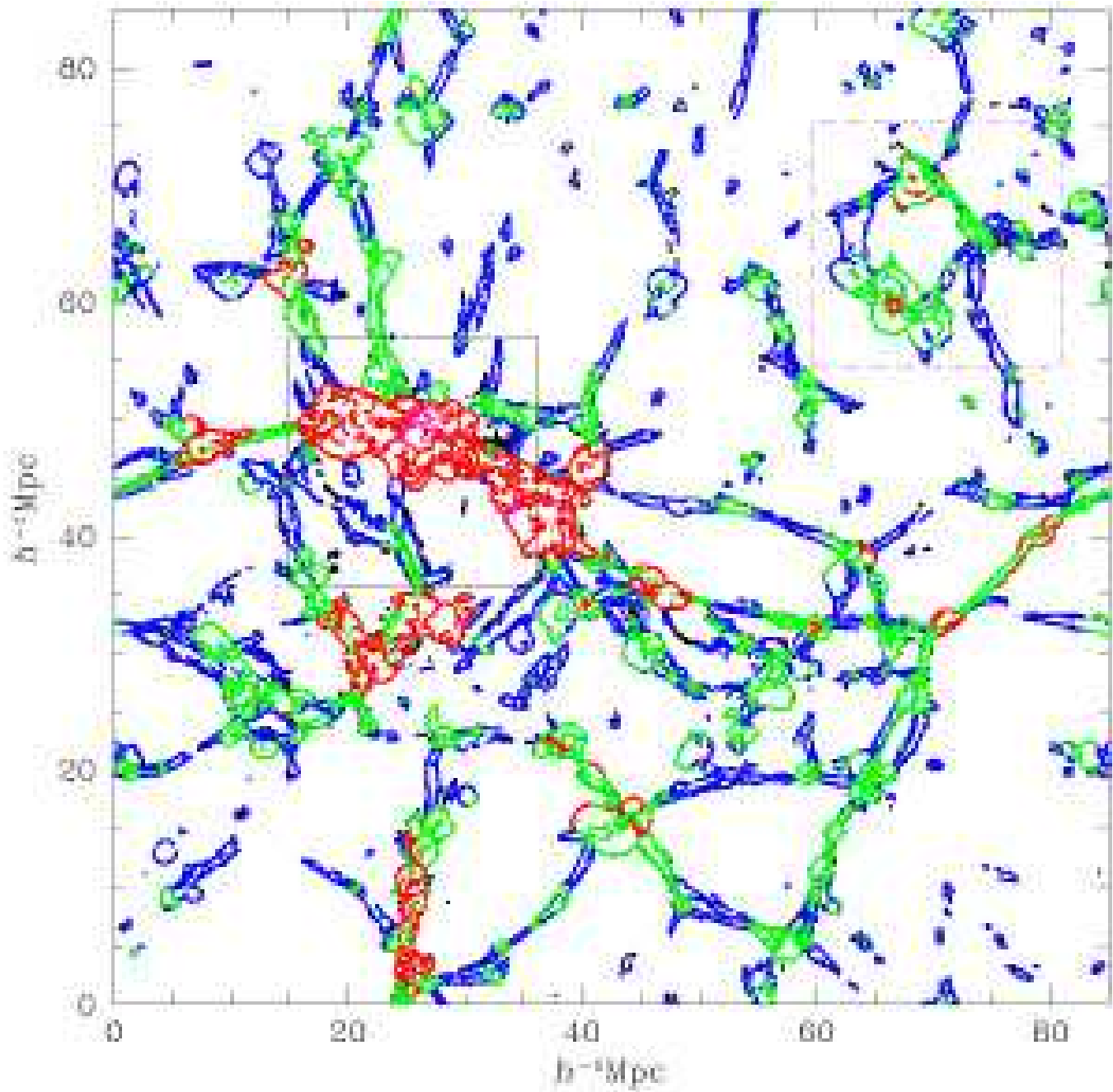}}
\figcaption{Two-dimensional slice of $(85\Mpc)^2$ showing shock
locations at $z=0$ in the GSW simulation, which are color-coded
according to shock speed as follows:
black for $V_s< 15\kms$, blue for $15 \le V_s< 65\kms$, 
green for $65 \le V_s< 250\kms$, red for $250 \le V_s< 1000\kms$, 
and magenta for $V_s \ge 1000\kms$. 
A blown-up image of the box (dashed line) in the upper right corner
is shown in Figure 3, while a blown-up image of the box (solid line)
around two merging clusters is shown in Figure 7.}
\end{figure}

\clearpage

\begin{figure}
\vspace{-1cm}
\centerline{\epsfxsize=18cm\epsfbox{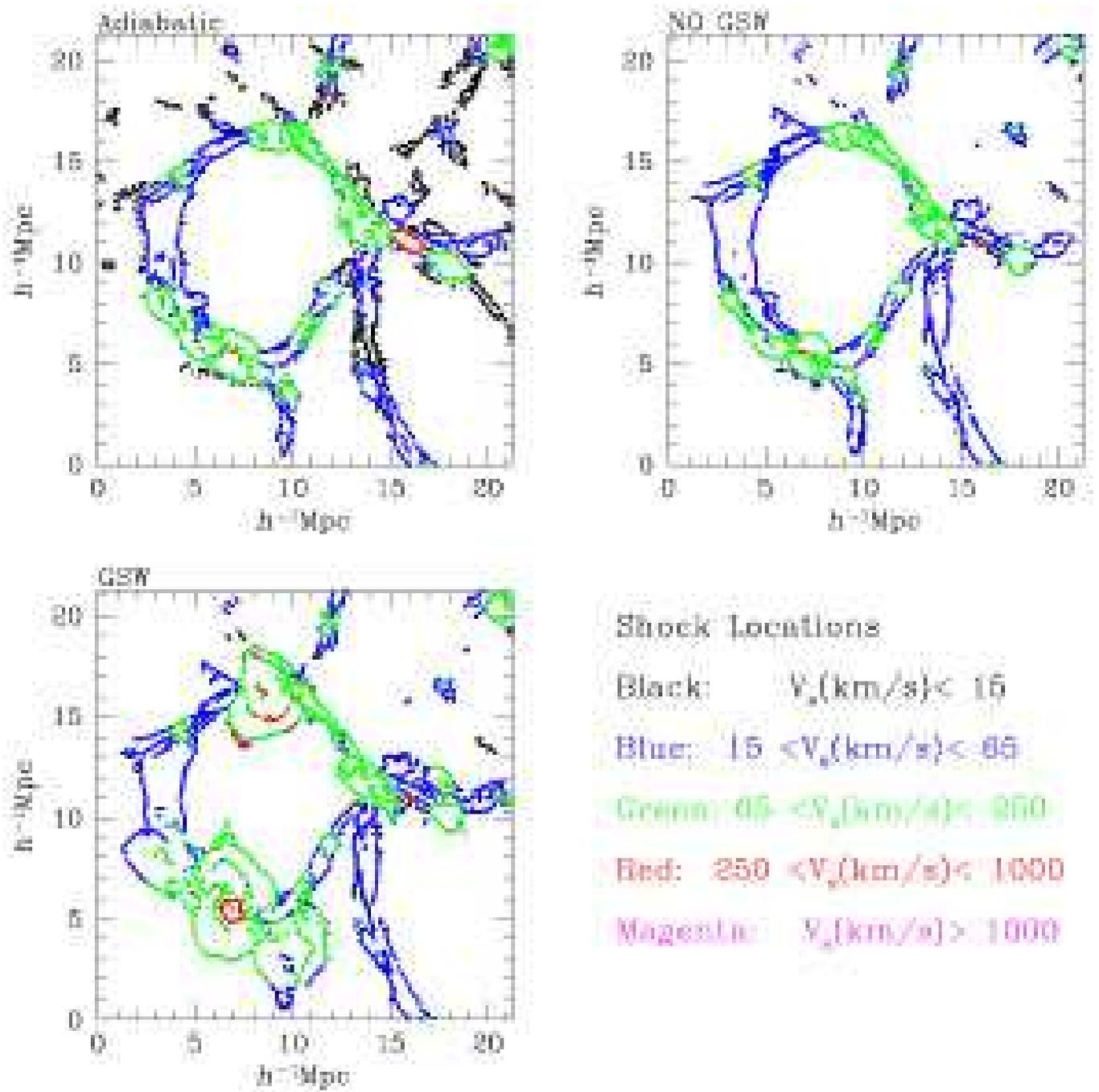}}
\figcaption{Two-dimensional slice of $(21.25\Mpc)^2$ showing shock
locations at $z=0$ in the Adiabatic, NO GSW and GSW simulations.
The locations are color-coded according to shock speed.
Two groups in the GSW simulation have $kT\sim 0.2-0.3$ keV.}
\end{figure}

\clearpage

\begin{figure}
\vspace{-1cm}
\centerline{\epsfxsize=18cm\epsfbox{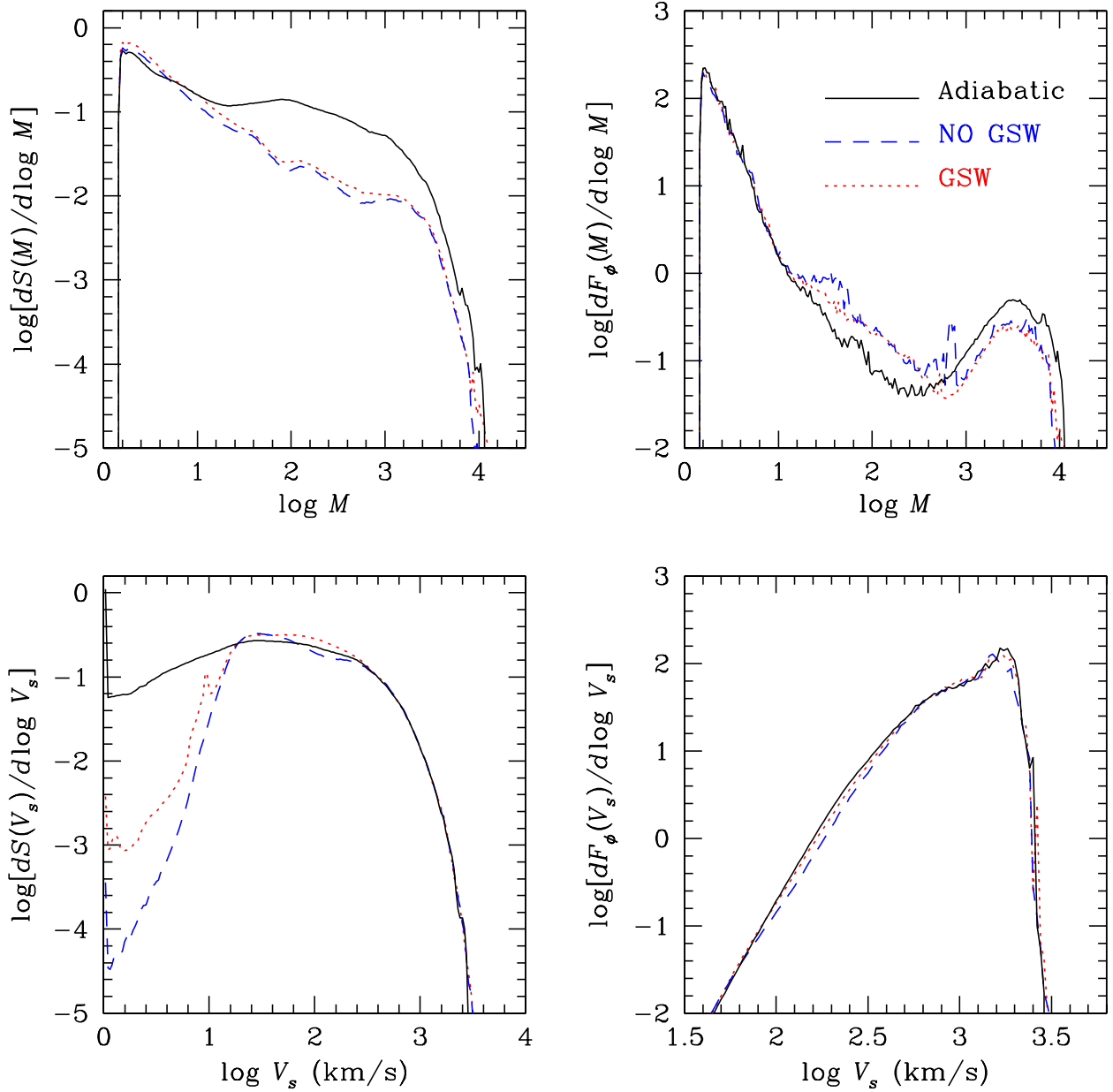}}
\figcaption{{\it Left panels:} Inverse of the mean comoving distance
between shock surfaces as a function of Mach number $M$ (top) and
shock speed $V_s$ (bottom) at $z=0$ for the Adiabatic (solid line),
NO GSW (dashed line), and GSW (dotted line) simulations. 
{\it Right panels:} Kinetic energy flux per comoving volume
passing through shock surfaces in units of
$10^{40}$ ergs s$^{-1}$ $(h^{-1}{\rm Mpc})^{-3}$ as a function of
$M$ (top) and $V_s$ (bottom).
Note that the bottom two panels have different ranges of abscissa.}
\end{figure}

\clearpage

\begin{figure}
\vspace{-1cm}
\centerline{\epsfxsize=18cm\epsfbox{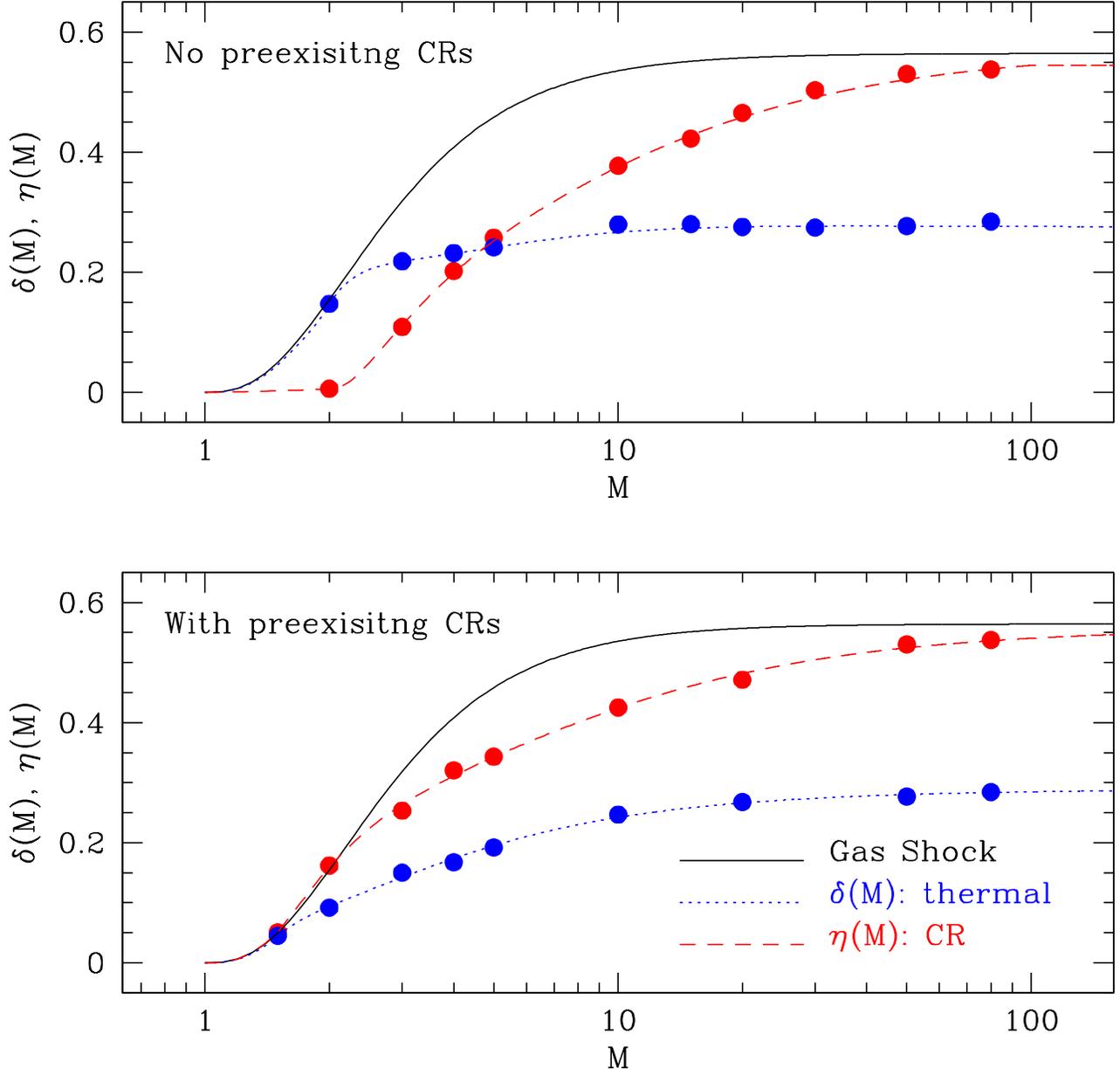}}
\figcaption{Gas thermalization efficiency, $\delta(M)$, and CR
acceleration efficiency, $\eta(M)$, as a function of Mach number.
Red and blue dots are the values estimated from numerical simulations
based on a DSA model and red and blue lines are the fits.
The top panel shows the case without preexisting CRs, while the bottom
panel shows the case with preexisting CRs at a level of
$P_{CR}/P_g \sim 0.3$ in the preshock region.
Black solid line is for the gas thermalization efficiency for shocks
without CRs.}
\end{figure}

\clearpage
\begin{figure}
\vspace{-1cm}
\centerline{\epsfxsize=17cm\epsfbox{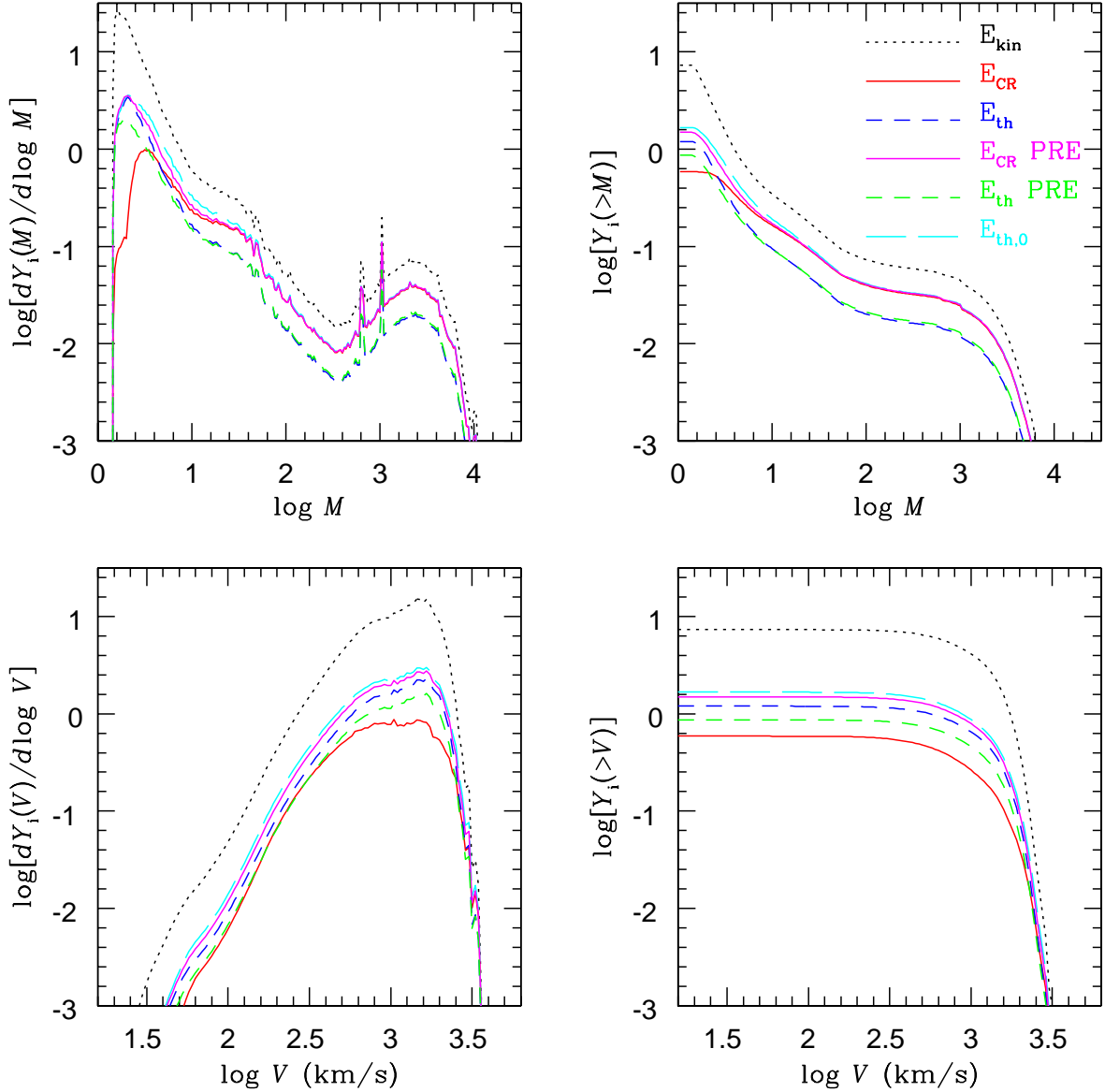}}
\vspace{-1cm}
\figcaption{{\it Left panels}: Shock kinetic energy passed, $dY_{\phi}$
(dotted line), thermal energy dissipated, $dY_{th}$ (dashed line), and
CR energy dissipated, $dY_{CR}$ (solid line), through surfaces of
cosmological shocks with Mach number between $\log M$ and
$\log M + d (\log M)$ (top) and through surfaces of cosmological
shocks with shock speed between $\log V_s$ and $\log V_s + d (\log V_s)$
(bottom), integrated from $z=5$ to $z=0$.
Red and magenta lines are the CR energy for the cases without and with
preexisting CRs, respectively.
Blue and green lines are the thermal energy for the cases without and
with preexisting CRs, respectively.
The thermal energy expected to be dissipated at cosmological shocks
without CRs (long dashed cyan line) is also plotted for comparison.
{\it Right panels}: Cumulative energy distributions, $Y_i(>M)$ (top)
and  $Y_i(>V_s)$ (bottom), for Mach number greater than $M$ and
for shock speed greater than $V_s$. 
The energies are normalized by the gas thermal energy  at $z=0$
inside the simulation box}
\end{figure}

\clearpage

\begin{figure}
\vspace{-1cm}
\centerline{\epsfxsize=18cm\epsfbox{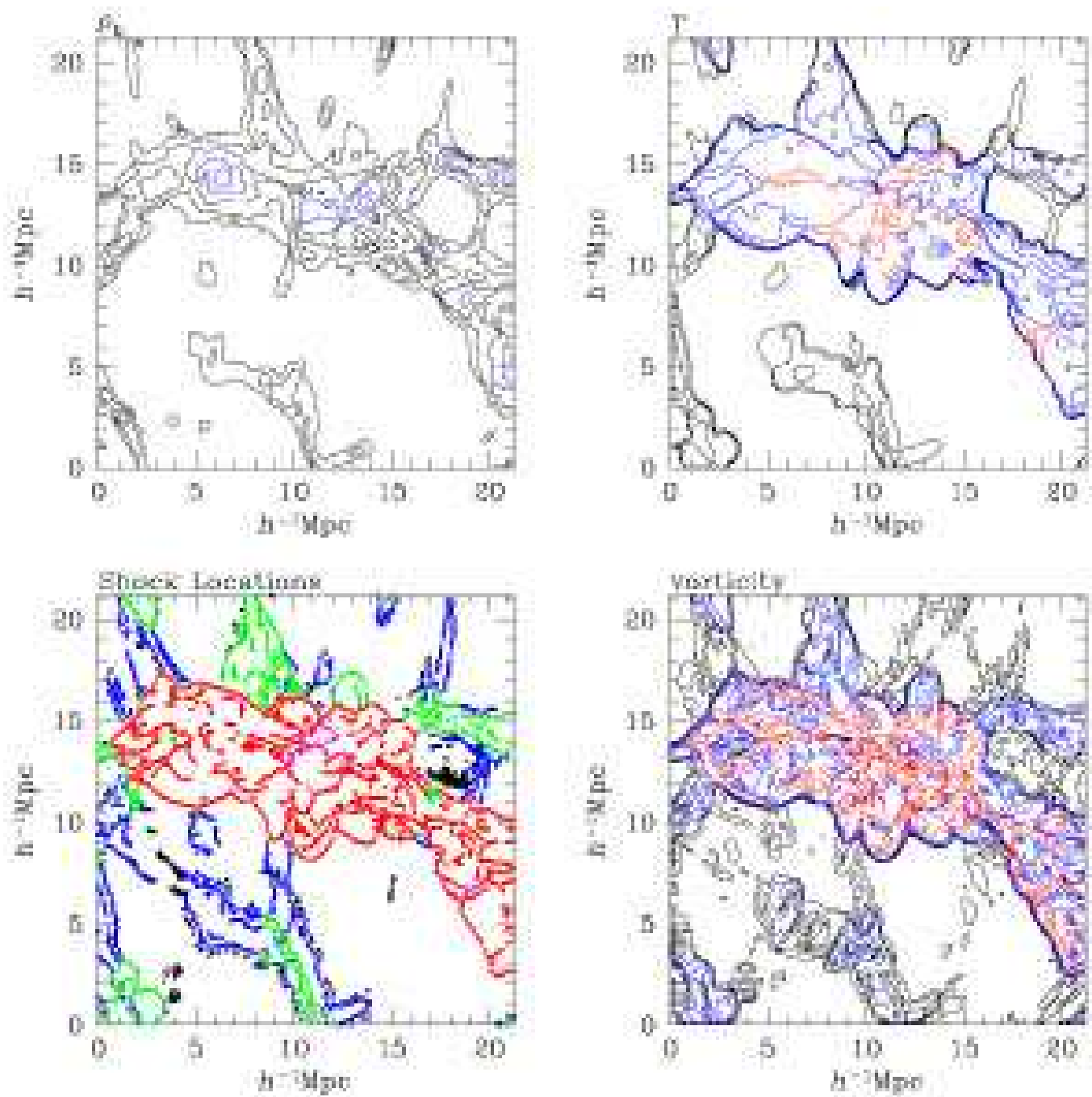}}
\figcaption{Two-dimensional slice of $(21.25\Mpc)^2$ around two merging
clusters with $kT \sim 1-2$ keV at $z=0$ in the GSW simulation.
Distributions of gas density (top left), temperature (top right), shock
locations (bottom left), and vorticity (bottom right) are shown.
In the gas density, temperature, and vorticity distributions, back, blue
and red contours represent regions of low, middle, and high values,
respectively.}
\end{figure}

\clearpage

\begin{figure}
\vspace{-1cm}
\centerline{\epsfxsize=18cm\epsfbox{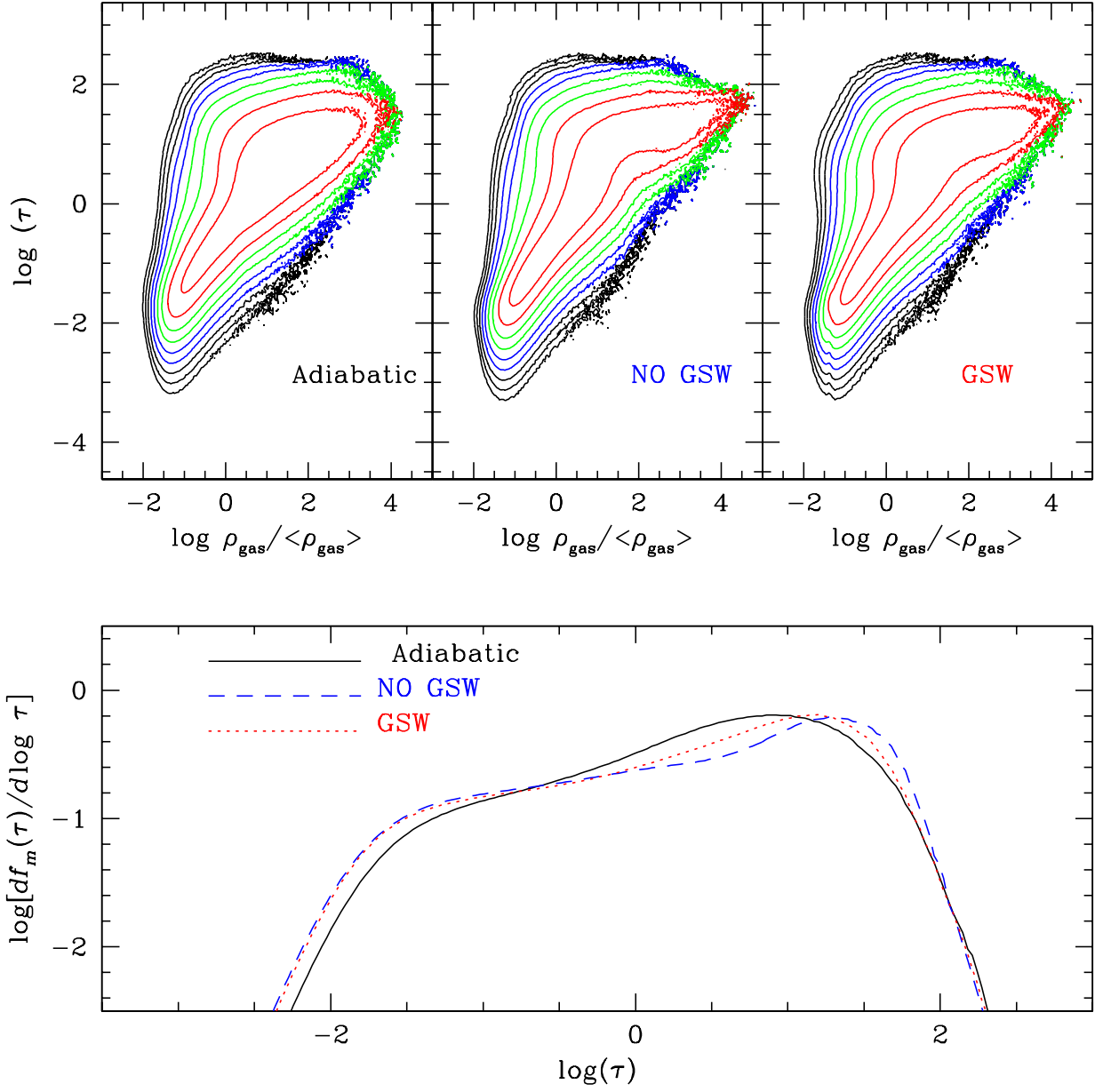}}
\figcaption{{\it Top panels:} Gas mass distribution in the gas mass
density-vorticity parameter plane at $z=0$ for the Adiabatic, NO GSW,
and GSW simulations.
The vorticity parameter is defined as $\tau = \omega t_{age}(z)$, 
where $\omega = |{\vec \nabla} \times {\vec v} |$ and $t_{age}(z)$ is
the age of the universe at redshift $z$.
{\it Bottom panel:} Gas mass fraction  as a function of vorticity
parameter at $z=0$ for the three simulations.}
\end{figure}

\end{document}